\documentclass[sigconf]{acmart}

\AtBeginDocument{%
  }

\setcopyright{acmlicensed}
\copyrightyear{2025}
\acmYear{2025}
\acmConference[LLM-DPM 2025]{LLM-DPM 2025 workshop (in conjunction with SIGMOD/PODS 2025)}{2025}{Berlin, Germany}

\usepackage{subfigure} 

\begin{document}

\title{Text2Cypher: Data Pruning using Hard Example Selection}

\author{Makbule Gulcin Ozsoy}
\email{makbule.ozsoy@neo4j.com}
\orcid{0000-0001-6013-1668}
\affiliation{%
  \institution{Neo4j}
  \city{London}
  \country{UK}
}

\thanks{Accepted to LLM-DPM 2025 workshop (in conjunction with SIGMOD/PODS 2025)}

\renewcommand{\shortauthors}{Ozsoy et al.}

\begin{abstract}

Database query languages such as SQL for relational databases and Cypher for graph databases have been widely adopted. Recent advancements in large language models (LLMs) enable natural language interactions with databases through models like Text2SQL and Text2Cypher. Fine-tuning these models typically requires large, diverse datasets containing non-trivial examples. However, as dataset size increases, the cost of fine-tuning also rises. This makes smaller, high-quality datasets essential for reducing costs for the same or better performance. In this paper, we propose five hard-example selection techniques for pruning the Text2Cypher dataset, aiming to preserve or improve performance while reducing resource usage. Our results show that these hard-example selection approaches can halve training time and costs with minimal impact on performance, and demonstrates that hard-example selection provides a cost-effective solution. 
\end{abstract}






\maketitle

\section{Introduction}
In today's world, data and knowledge are stored, managed, and queried through databases,  which are accessed using query languages such as SQL (for relational databases) or Cypher (for graph databases). Recent advancements in large language models (LLMs) have made it possible to interact with databases using natural language, allowing models like Text2SQL and Text2Cypher to translate natural language questions into database queries. A common approach for generating these queries is to fine-tune foundational models using question-query datasets. 
Effective fine-tuning of these models requires large, diverse datasets with non-trivial examples.

\begin{figure}
    \centering
    \includegraphics[width=0.95\linewidth]{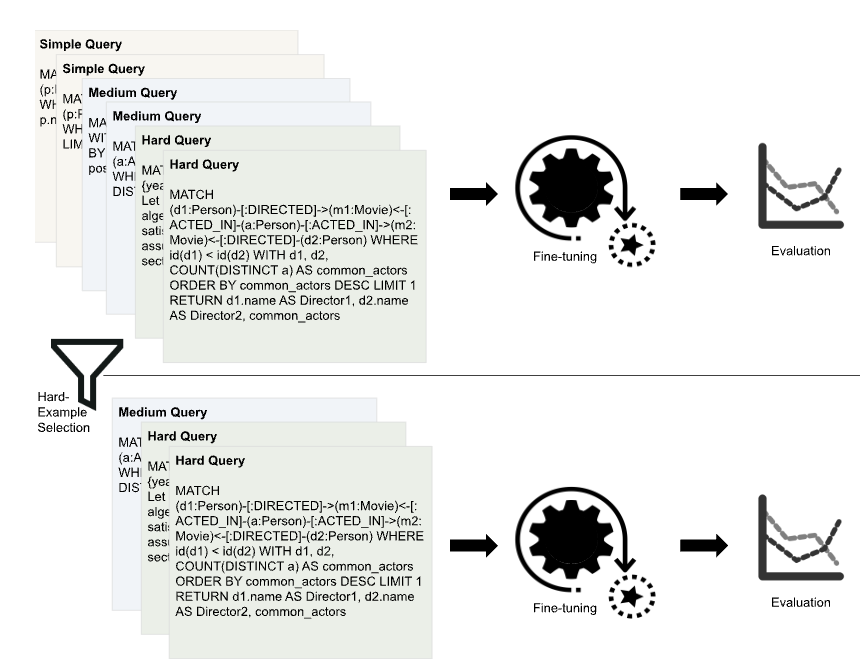}    
    \caption{Hard-Example Selection for Dataset Pruning}
    \Description{Hard-Example Selection for Dataset Pruning}
    \label{fig:overall_procedure}
\end{figure}

With increased use of synthetic datasets, it is now possible to automatically generate larger datasets. However, these datasets often suffer from quality and redundancy issues. Recent research suggests that small, high-quality datasets can outperform larger ones when fine-tuning LLMs \cite{zhou2023lima, zhang2025d3}. Additionally, the cost of fine-tuning LLMs increases as the dataset size grows. One way to address these challenges is to prune or select a subset of the data. This process should be automated to ensure that the resulting dataset (i) maintains high performance and (ii) minimizes costs, achieving greater efficiency \cite{lin2024data}. 
Figure \ref{fig:overall_procedure} shows a hard-example selection procedure. Initially, we start with a larger dataset containing simple, medium, and hard Cypher queries used for fine-tuning a Text2Cypher model. After applying hard-example selection, the dataset is reduced in size and predominantly retains medium and hard queries.

In this paper, we apply five hard-example selection approaches to prune the Text2Cypher dataset: three approaches for selecting challenging instances from a larger training dataset to enhance model performance and two approaches that combine the proposed hard-example selection methods. 
We evaluate their impact on a Text2Cypher dataset, analyzing training time (in terms of training steps) and Cypher generation performance. Our main contributions are: 
\begin{itemize}
\item We propose hard-example selection techniques specifically for the Text2Cypher task. Three approaches leverage prior analysis results and heuristics to identify challenging (hard) examples and prune the training dataset, while two additional approaches combine these methods to improve performance.
\item We analyze their impact on the Text2Cypher task on training time (measured in steps), loss values, and Cypher generation performance.
\item Our results show that hard-example selection approaches reduce resource usage — both in elapsed time and total cost— by more than half while minimally affecting Cypher generation performance. Although there is room for improvement in matching the performance of training on the full dataset, hard-example selection presents a cost-effective solution.
\end{itemize}

The structure of the paper is as follows: Section \ref{rel_work} reviews related work on data subset selection and pruning, particularly for fine-tuning large language models. Section \ref{hard-example-selection} details the hard-example selection approaches applied to the Text2Cypher task. Section \ref{experiment} outlines our experimental setup and presents the evaluation results. Finally, Section \ref{conc} provides the conclusion.

\section{Related work} \label{rel_work}

Several approaches for data selection or pruning have been proposed in the literature \cite{wang2024survey, albalak2024survey}, ranging from the use of baseline LLM models to decide which instances to select or create embeddings \cite{chen2024alpagasus, liu2024makes, chen2023maybe, du2023mods}, to methods that rely on instance-level scores based on system indicators like diversity or difficulty. 
For example, Maharana et al. \cite{maharana2024d2} use graph-based techniques to reduce redundancy by iteratively selecting diverse and challenging instances.
Lin et al. \cite{lin2024data} utilize influence and effort scores to prioritize influential and difficult samples for fine-tuning. 
Zhang et al. \cite{zhang2025d3} identify diverse, difficult, and dependable data iteratively. In each iteration, they evaluate the distinctiveness, difficulty (through uncertainty-based prediction), and dependability (using an external LLM) of instances, then apply a weighted function to select a subset. 
Tan et al. \cite{tan2025data} propose InfoMax, selecting samples based on informativeness and overlap between pairwise samples. 

Other approaches include training a model on a small subset, then using it to prune the data. 
For example, Li et al. \cite{li2024quantity} fine-tune a model on a randomly sampled subset of data, then use the fine-tuned model to calculate Instruction Following Difficulty (IFD) scores for each instance. Instances with greater difficulty, based on IFD score, are selected for final fine-tuning. 
Xu et al. \cite{xu2023hard} focuses on differentiating informative hard samples from misleading ones in model training. In their HardPT framework, they utilize reinforcement learning and adaptive contrastive learning techniques.  
Azeemi et al. \cite{azeemi2023data} employ cross-entropy scores to select harder instances. In their experiments they observe that selecting more difficult instances results in improved model performance. 
Xia et al. \cite{xia2024less} introduce the LESS algorithm, an optimizer-aware approach for efficient data selection. It uses a warm-up training phase to generate low-dimensional gradient features, which are stored and later used by models for training. 
Finally, Yang et al. \cite{yang2025diversity} focus on diversity-aware selection using sparse autoencoders and either greedy-sampling approach (SAE-GreedSelect) or similarity-based sampling (SAE-SimScale) approach.

Although data selection or pruning are well-studied in machine learning, their application to natural language to query language tasks, such as Text2SQL and Text2Cypher, remains largely unexplored. 
SE-HCL \cite{zhang2024se} applies curriculum learning to the Text2SQL task by training the model progressively, starting with easy instances and gradually moving to more difficult ones. This approach involves iterative steps that begin with simplifying the data, gradually increasing its complexity, and evaluating the difficulty of individual instances. 
Some Text2SQL datasets, such as Spider \cite{yu2018spider} and IndDB \cite{nascimento2024text}, provide difficulty labels based on SQL constructs like GROUP BY clauses and nested subqueries, where more complex constructs indicate higher difficulty. However, these difficulty annotations are primarily used for analyzing evaluation outputs rather than for data selection. 
In this work, we explore data pruning for the Text2Cypher task by focusing on hard-example selection based on instance difficulty. 

\section{Hard-Example Selection for Text2Cypher}\label{hard-example-selection}
We introduce five methods for selecting hard examples. Three of them focus on finding more challenging instances, while the other two combine these approaches to improve selection.

\subsection{Selecting Challenging Instances}

In our previous work \cite{neo4jAnalyzingModelStruggles}, we have executed a comprehensive analysis of model performance on the Neo4j Text2Cypher (2024) dataset  \cite{ozsoy2025text2cypher}. 
This analysis explored evaluation results from multiple perspectives, including key metrics (such as Google-Bleu and Exact Match), assigned complexity levels, and breakdowns by data source, database type, and fine-tuned model. 
The statistical analyses (e.g., averages, standard deviations) revealed consistent patterns of model struggle, particularly on examples from specific databases and data sources. Further error analysis highlighted that these challenges were often due to inconsistencies in ground-truth Cypher queries (such as varying use of WHERE clauses or aggregation methods), limitations of existing evaluation metrics, and underlying model weaknesses. These findings directly motivated our hard-example selection strategies, which aim to construct a more informative and targeted training subset by focusing on the most challenging instances.

In this section, we describe three approaches for selecting challenging instances from a larger training dataset to enhance model performance. 

\begin{itemize}
    \item \textbf{Complexity-Based Hard-Example Selection}: Guided by our analysis \cite{neo4jAnalyzingModelStruggles}, we identified data sources and databases where fine-tuned models struggled most. Based on this analysis: (i) The chosen databases are three demonstration databases of Neo4j~\footnote{Neo4j Text2Cypher Crowdsourcing App: \url{https://text2cypher.vercel.app/}}~\footnote{Neo4j Browser Demo: \url{https://demo.neo4jlabs.com:7473/browser/}}, namely "recommendations, companies, neoflix", and (ii) The selected data-sources are: "functional\_cypher",  "synthetic\_gemini", and "text2cypher2023\_train". 
    
    For the selection of these instances, we used a logical "OR" to include instances from either the selected databases or data sources. While this results in a diverse set of challenging instances, we observe an imbalance with many instances coming from a single data source. To address this, we performed additional sampling, limiting each group to a maximum of 4,000 instances (the average group size). This resulted in a total of 16,173 instances, less than the half of the original training dataset, of approximately 40K instances.
    
    \item \textbf{Length-Based Hard-Example Selection}: This heuristic approach assumes that longer ground-truth Cypher queries are more challenging for a language model to generate owing to their increased complexity. Longer queries often involve multiple clauses, making them harder to replicate accurately. 
    Therefore, this approach selects instances based on the length of the Cypher query. To ensure consistency with other selection methods, we maintained a final dataset size of 16,173 instances.
     
    \item \textbf{Cypher-Specific Hard-Example Selection}: This heuristic method focuses on the presence of Cypher-specific terms (e.g., MATCH, WHERE, RETURN), under the assumption that queries containing more such terms are more complex. Unlike the length-based approach, which prioritizes the length of queries, this method selects instances based on the count of Cypher terms, i.e., which are likely to be more complex by containing multiple clauses. To ensure fairness with other hard-instance selection methods, we restricted this dataset to 16,173 instances.

\end{itemize}

\subsection{Combining Selection Methods}

We combined the proposed hard-example selection approaches as follows:
\begin{itemize}
    \item \textbf{Complexity-Based \& Length-Based Hard-Example Selection}: After selecting hard examples using the Complexity-Based approach, we took an additional step to further refine the selection process. Specifically, we sorted the chosen instances in descending order based on the length of the Cypher queries. This step follows the methodology of the Length-Based approach, which assumes that longer queries tend to be more complex and, therefore, more challenging for the model to generate. By prioritizing longer queries, we made sure that the final set of hard examples was both challenging and diverse in terms of complexity.

    \item \textbf{Complexity-Based \& Cypher-Specific Hard-Example Selection}: Similar to the previous combined approach, after selecting hard examples using the Complexity-Based approach, we ranked them by the number of Cypher-specific terms in descending order, aligning with the Cypher-Specific approach. This method emphasizes instances with more Cypher-specific terms, as these tend to be more complex and involve multiple clauses. The final subset, therefore, includes challenging instances and have a diverse set of complexities.

\end{itemize}

\subsection{Baseline Approaches}
We used the following baseline approaches: 
\begin{itemize}
    \item \textbf{Original Data}: This baseline uses the training data without any modifications which provides a reference point for performance comparisons.
    
    \item \textbf{Randomly-Sampled}: In this approach, we randomly sampled instances from the original data. To ensure fairness with the Complexity-Based approach, we aimed to create a balanced dataset across data source groups. We first sampled each group (based on the data-source field) to a size of 2,755, representing the 75th percentile of data source group sizes. We then refined the sample to 16,173 instances to match the size used in the hard-instance selection methods.

\end{itemize}

\section{Experimental Setup and Results }\label{experiment}

\subsection{Experimental Setup and Evaluation Metrics}
For our experiments, we used the publicly available Text2Cypher dataset \cite{ozsoy2025text2cypher}, which contains 44,387 instances—39,554 for training and 4,833 for testing. This dataset is a cleaned and combined version of multiple data sources, most of which were synthetically generated.

We employed two evaluation procedures to measure model performance:
(i) Translation-Based (Lexical) Evaluation: This method compares generated Cypher queries with ground-truth queries at the textual level.
(ii) Execution-Based Evaluation: This method executes both the generated and ground-truth Cypher queries on the target database and compares their outputs, sorted lexicographically. This approach requires an active target database, where about 50\% of the dataset has such references. As a result, it evaluates only a subset of the data. 
To compute these evaluation metrics, we used the Hugging Face Evaluate library \cite{hfEvaluate}. We report the Google-Bleu and Exact Match scores as the primary evaluation metrics.

We fine-tuned a baseline model, 'unsloth/Meta-Llama-3.1-8B-Instruct-bnb-4bit', using various training datasets prepared according to the proposed hard-example selection methods. 
During evaluation, we used the test set and fine-tuned models to generate Cypher queries based on input natural language questions and corresponding database schemas. After generating the Cypher queries, we applied a post-processing step to remove unwanted text, such as the 'cypher:' prefix. Details of the fine-tuning setup and parameters are provided in Appendix \ref{finetune_parameters}.

\subsection{Evaluation Results}
We analyzed the impact of (i) using a subset of the full dataset, assessing both training efficiency and model accuracy, and (ii) applying different hard-example selection approaches on performance.

\begin{figure}
    \centering
    \subfigure[Training loss: Original vs. Randomly-Sampled data]{
        \includegraphics[width=0.95\linewidth]{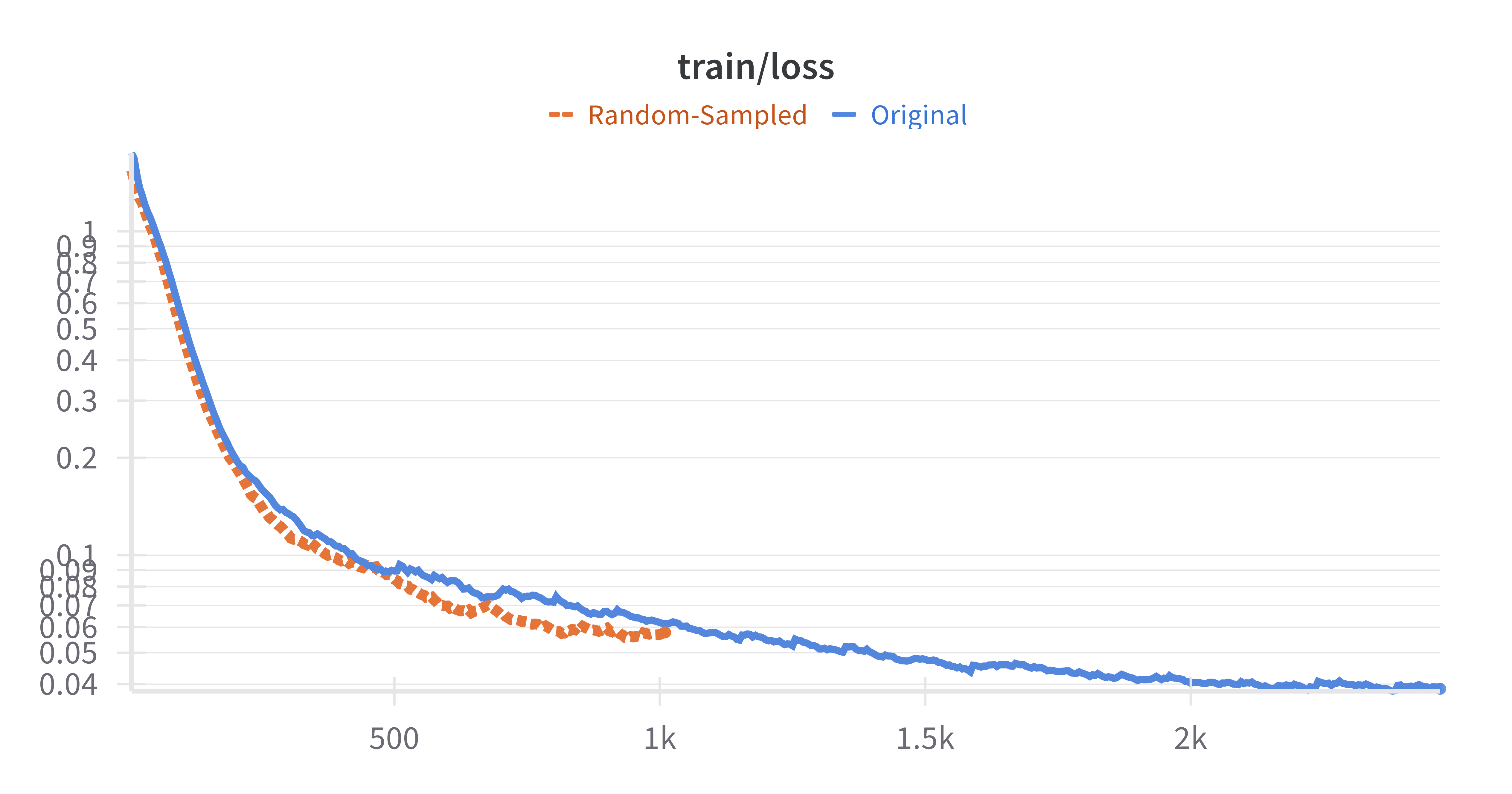}
    }
    \subfigure[Translation-based - Google-Bleu score]{
        \includegraphics[width=0.47\linewidth]{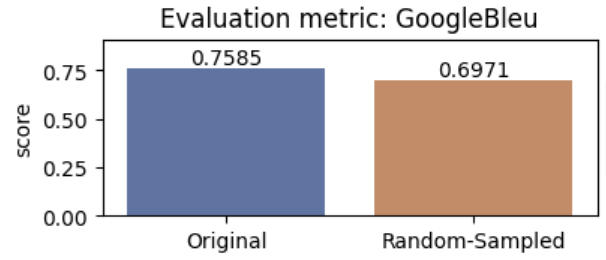}
    }
    \subfigure[Translation-based - Exact-Match score]{
        \includegraphics[width=0.47\linewidth]{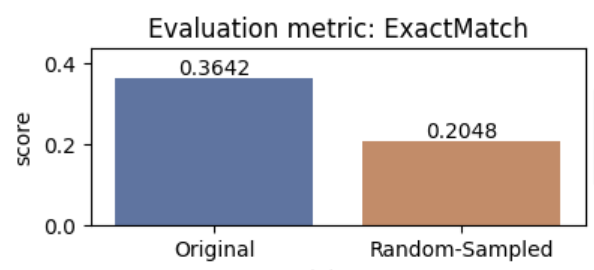}
    }
     \subfigure[Execution-based - Google-Bleu score]{
        \includegraphics[width=0.47\linewidth]{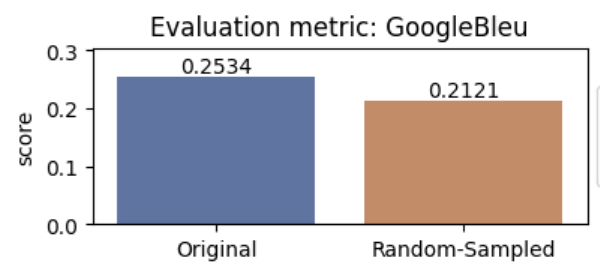}
    }
    \subfigure[Execution-based - Exact-Match score]{
        \includegraphics[width=0.47\linewidth]{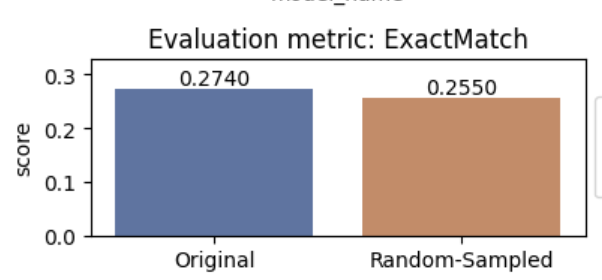}
    }
    
    \caption{Original vs. Randomly-Sampled data}
    \Description{Original vs. Randomly-Sampled data}
    \label{fig:eval_org_vs_random}
\end{figure}

\subsubsection{Impact of Training Data Reduction}
The original 40K-instance training dataset was reduced to 16,173 instances through random-sampling or hard-example selection. As shown in Figure \ref{fig:eval_org_vs_random}, training the full dataset required around 2.5K steps (batch size 16), while the subset datasets needed only 1K steps. This reduction significantly cut fine-tuning time and costs. Using subset data achieved comparable or better training loss at 1K steps. However, over the full 2.5K steps, the original full dataset achieved a better final loss: 0.0387 versus 0.0569 for random sampling. 
Translation-based evaluation, which is based on token prediction accuracy, aligns closely with the loss function. The original dataset achieved a Google-Bleu score of 0.75 and an Exact Match score of 0.36, whereas the random sampling approach scored lower at 0.69 and 0.20, respectively. Execution-based evaluation showed smaller drops, with the full dataset scoring 0.25 (Google-Bleu) and 0.27 (Exact Match) versus 0.21 and 0.25 for the randomly sampled dataset. 
In summary, using subsets cuts training time and costs by over half but reduces performance. We next explore whether hard-example selection can retain efficiency while improving outcomes.

\begin{figure}
    \centering
    \subfigure[Training loss: Randomly-Sampled and Hard-Example Selection approaches]{
        \includegraphics[width=0.95\linewidth]{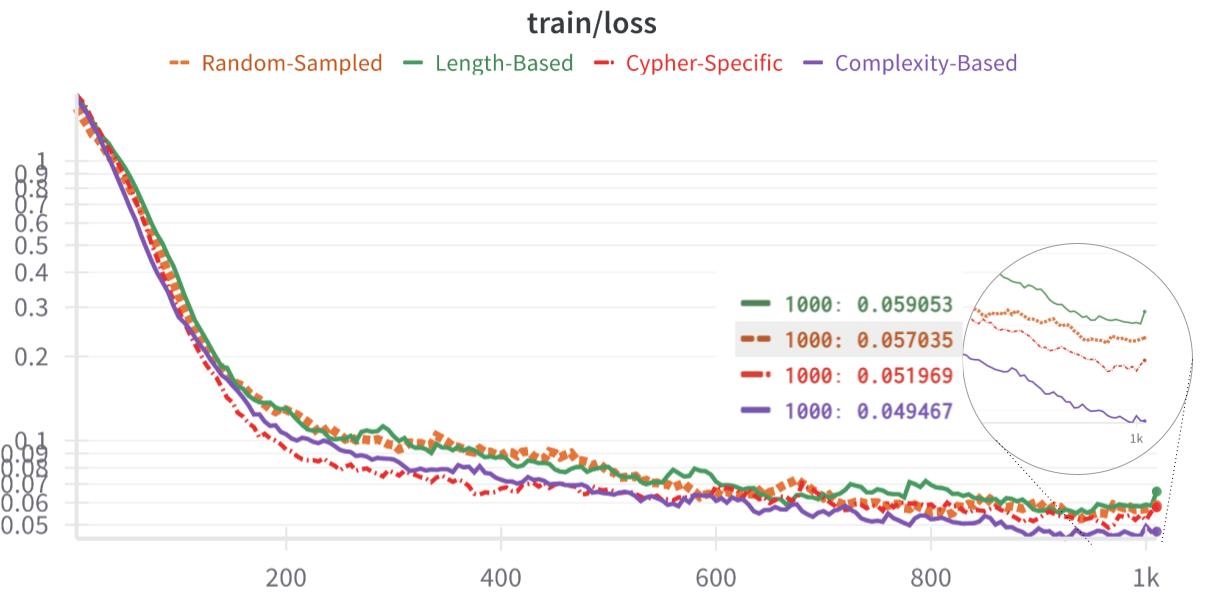}
    }
    \subfigure[Translation-based - Google-Bleu score]{
        \includegraphics[width=0.47\linewidth]{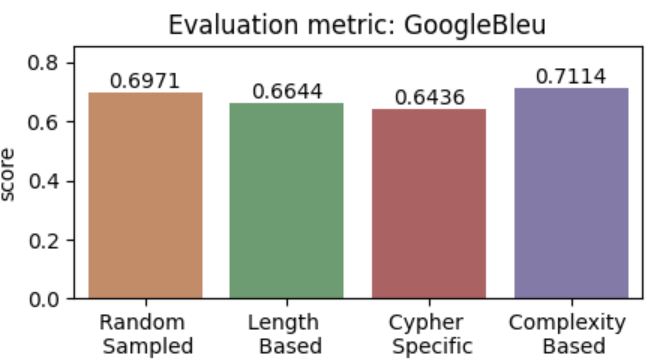}
    }
    \subfigure[Translation-based - Exact-Match score]{
        \includegraphics[width=0.47\linewidth]{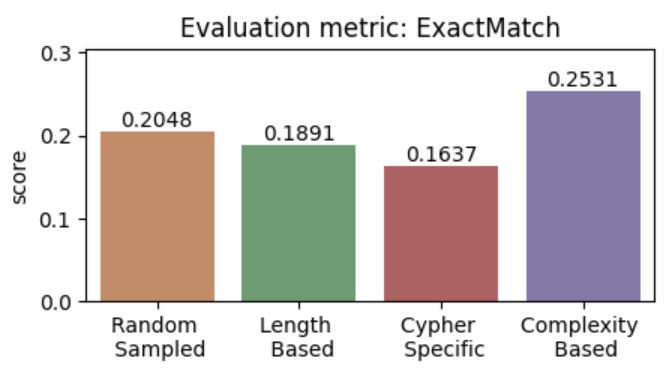}
    }
     \subfigure[Execution-based - Google-Bleu score]{
        \includegraphics[width=0.47\linewidth]{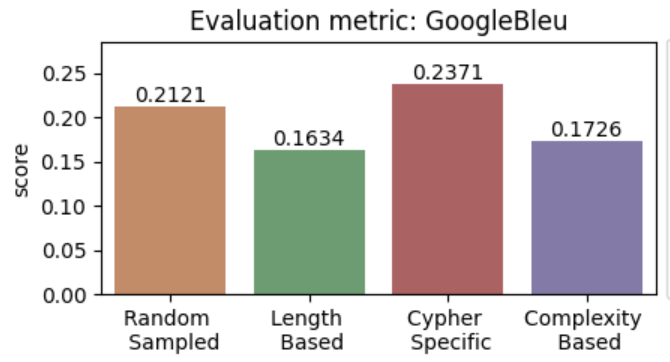}
    }
    \subfigure[Execution-based - Exact-Match score]{
        \includegraphics[width=0.47\linewidth]{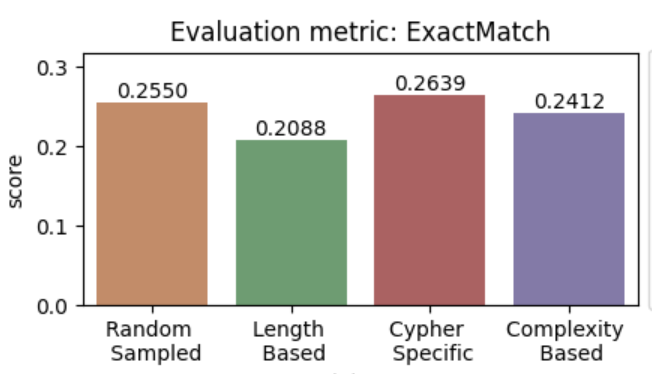}
    }
    
    \caption{Randomly-Sampled and Hard-Example Selection approaches}
    \Description{Randomly-Sampled and Hard-Example Selection approaches}
    \label{fig:eval_hard_examples}
\end{figure}

\subsubsection{Impact of Hard-Example Selection}
When fine-tuning the baseline model with datasets prepared using random sampling or hard-example selection approaches, training times remain similar since the dataset sizes were kept equal, as shown in Figure \ref{fig:eval_hard_examples}. All methods achieve comparable loss values, ranging between 0.05 and 0.06. However, closer inspection reveals a ranking from highest (worst) to lowest (best) loss: Length-Based → Random-Sampled → Cypher-Specific → Complexity-Based. In translation-based evaluation, the Complexity-Based approach performs best, achieving 0.71 Google-Bleu and 0.25 Exact Match, bringing it closer to the performance of the original dataset. Interestingly, execution-based evaluation, which is run on a subset of data that has access to active demonstration databases, follows a different pattern. In this case, the Cypher-Specific approach yields the best results, with Google-Bleu and Exact Match scores of 0.23 and 0.26, respectively. 


\begin{figure}
    \centering
    \subfigure[Training loss: Combined Hard-Example Selection approaches]{
        \includegraphics[width=0.95\linewidth]{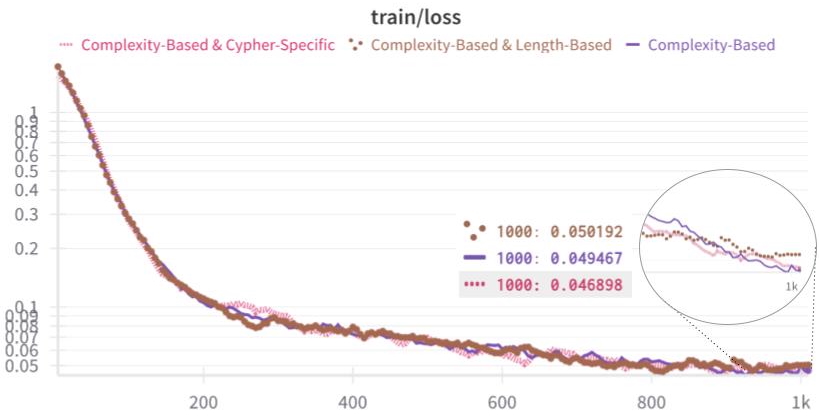}
    }
    \subfigure[Translation-based - Google-Bleu score]{
        \includegraphics[width=0.47\linewidth]{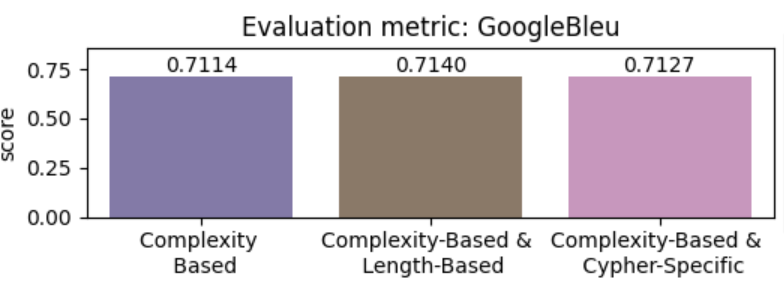}
    }
    \subfigure[Translation-based - Exact-Match score]{
        \includegraphics[width=0.47\linewidth]{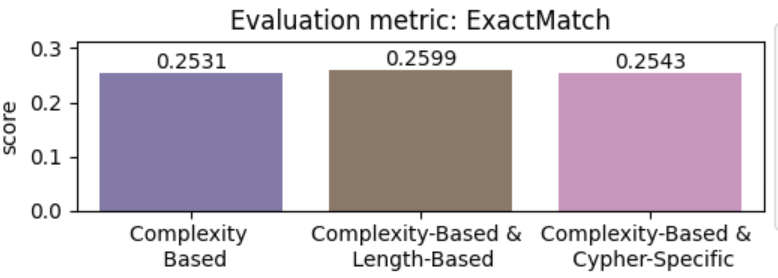}
    }
     \subfigure[Execution-based - Google-Bleu score]{
        \includegraphics[width=0.47\linewidth]{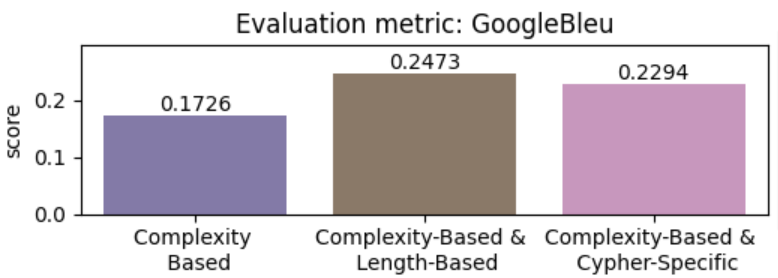}
    }
    \subfigure[Execution-based - Exact-Match score]{
        \includegraphics[width=0.47\linewidth]{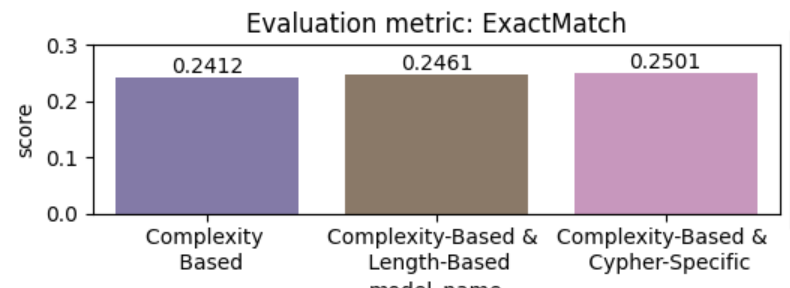}
    }
    
    \caption{Combined Hard-Example Selection approaches}
    \Description{Combined Hard-Example Selection approaches}
    \label{fig:eval_combined}
\end{figure}

\subsubsection{Impact of Combining Approaches on Performance}
Combining the Complexity-Based approach with either the Length-Based or Cypher-Specific approach did not result in significantly different loss values, as shown in Figure \ref{fig:eval_combined}. For translation-based evaluation, all approaches performed similarly, with Google-Bleu and Exact Match scores around 0.71 and 0.25, respectively. However, execution-based evaluation revealed some variation: The best Google-Bleu score (0.24) is achieved by Complexity-Based \& Length-Based approach, and the best Exact Match score (0.25) is achieved by Complexity-Based \& Cypher-Specific approach. 
These findings suggest that although combining approaches does not drastically impact performance, some combinations may offer slight advantages depending on the evaluation method.

\subsubsection{Overall}
As shown in Table \ref{tab:overall_comparison}, while the full dataset achieves the highest Google-Bleu and Exact Match scores for both translation- and execution-based evaluation, hard-example selection outperforms random sampling. It also reduces resource usage—time and cost—by more than half, as presented in Figure \ref{fig:eval_org_vs_random}, with minimal performance loss. 
We observe that fine-tuned models may still benefit from more data or better-tuned hyper-parameters, even with 16K instances. Future work will explore increasing data diversity and optimizing hyper-parameters to boost performance. Additionally, the difference between evaluation methods requires further investigation. While translation-based evaluation closely aligns with the loss function, reflecting token prediction accuracy, execution-based evaluation follows a different pattern. We attribute this behavior to the fact that execution-based evaluation is run on instances that have access to demonstration databases, which is around 50\% of the dataset. In the future, we will analyze how different data subsets impact the model's ability to generate accurate Cypher queries during execution-based evaluation.

\begin{table}
  \caption{Performance Comparison: Original (2.5K steps) vs. Randomly-Sampled (1K steps) vs. Hard-Example Selection (best scores - 1K steps)}
  \label{tab:overall_comparison}
  \begin{tabular}{p{0.25\linewidth}p{0.13\linewidth}p{0.13\linewidth}p{0.13\linewidth}p{0.13\linewidth}}
    \toprule
      & \multicolumn{2}{c}{\textbf{Translation-Based}} & \multicolumn{2}{c}{\textbf{Execution-Based}} \\
      & \textbf{Google-Bleu} &  \textbf{Exact-Match} &  \textbf{Google-Bleu} &   \textbf{Exact-Match}\\
    \midrule
    \textbf{Original} &  0.7585 & 0.3642 &  0.2534 & 0.2740 \\
     \textbf{Randomly\-Sampled} &  0.6971 &  0.2048 &  0.2121 &  0.2550 \\
     \textbf{Hard Example \newline Selection (best)} &  0.7140 &  0.2599  &  0.2473  &  0.2639\\
  \bottomrule
\end{tabular}
\end{table}

\section{Conclusion} \label{conc}
With models like Text2SQL and Text2Cypher, which translate natural language questions into database queries, it is now possible to interact with databases through natural language. In order to achieve this, foundational large language models (LLMs) are fine-tuned using large, diverse datasets containing non-trivial examples. However, the cost of fine-tuning these models can be significant, making it desirable to use smaller, high-quality datasets to optimize performance and resource usage. In this work, we explored hard-example selection for the Text2Cypher task, presenting five approaches to prune the training dataset. Our analysis demonstrates that selecting more complex or hard examples reduces resource usage, in terms of time and cost, by over half, while minimally affecting Cypher generation performance. This finding highlights the potential for smaller, high-quality datasets to optimize fine-tuning of large language models (LLMs), especially as a cost-effective strategy.

In this work, we focused on pruning the dataset by selecting the more complex (hard) instances. However, diversity of the data is also an important factor. In future research, we plan to explore pruning strategies that take both difficulty and diversity into account. We also aim to analyze how different subsets of the data affect the model’s ability to generate accurate Cypher queries and improve the dataset based on that. While we mostly used heuristic-based methods in this study, we plan to investigate more advanced techniques in the future, such as different loss functions and training strategies, to further boost model performance.

\bibliographystyle{ACM-Reference-Format}
\bibliography{main-arxiv}

\appendix

\section{Declaration on Generative AI Usage}

  
During the preparation of this work, the author(s) used Chat-GPT in order to: 'Improve writing style' and 'Paraphrase and reword'. After using these tool(s)/service(s), the author(s) reviewed and edited the content as needed and take(s) full responsibility for the publication’s content. 

\section{Fine-tuning Parameters} \label{finetune_parameters}

For fine-tuning, we used a RunPod \cite{runpod} GPU environment with a single A40 machine. The fine-tuning process was conducted using the Unsloth\cite{unsloth} framework. 
The parameters used for fine-tuning are presented in Table \ref{tab:finetune_parameters}. 

\begin{table}[hb!]
  \caption{Fine-tuning Parameters}
  \label{tab:finetune_parameters}
  \begin{tabular}{p{0.20\linewidth}|p{0.55\linewidth}}
    \toprule
     \textbf{Model \& \newline Tokenizer \newline Parameters} & 
        $max\_seq\_length: 2048$, \newline
        $dtype: torch.bfloat16$, \newline
        $load\_in\_4bit: True$, \newline
        $truncation\_side: "left"$, \newline
        $padding\_side: "left"$
    \\
    \midrule
    \textbf{PEFT \newline Parameters} &  
        $r: 8$, \newline
        $target\_modules: \newline [
            "q\_proj",
            "k\_proj",
            "v\_proj",
            "o\_proj"
        ]$, \newline
        $lora\_alpha: 16$, \newline
        $lora\_dropout: 0$, \newline
        $bias: "none"$, \newline
        $random\_state: 3407$, \newline
        $use\_rslora: False$, \newline
        $loftq\_config: None$
     \\
    \midrule
    \textbf{Training \newline Arguments} &   
        $per\_device\_train\_batch\_size: 2$,\newline
        $gradient\_accumulation\_steps: 4$,\newline
        $warmup\_steps: 5$,\newline
        $num\_train\_epochs: 1$,\newline
        $learning\_rate: 2e-4$,\newline
        $fp16: not is\_bfloat16\_supported()$,\newline
        $bf16: is\_bfloat16\_supported()$,\newline
        $optim: "adamw\_8bit"$,\newline
        $weight\_decay: 0.01$,\newline
        $lr\_scheduler\_type: "linear"$,\newline
        $seed: 3407$
        
     \\
  \bottomrule
\end{tabular}
\end{table}

\end{document}